\documentstyle[sprocl,epsfig,psfig]{article}

\def\be{\begin{equation}}
\def\ee{\end{equation}}
\def\bea{\begin{eqnarray}}
\def\eea{\end{eqnarray}}
\begin{document}

\title{NEW DEVELOPMENTS IN HYBRID PHOTON DETECTORS\footnote{
Presented at the conference
NEW DETECTORS, 
36th WORKSHOP of the INFN ELOISATRON PROJECT,
ETTORE MAJORANA CENTRE FOR SCIENTIFIC CULTURE,
Erice, Trapani, Sicily,
November 1-7, 1997. To appear in the proceedings,
World Scientific, Editors 
C. Williams and T. Ypsilantis.
}}

\author{DANIEL FERENC}

\address{Div. PPE,\\CERN,\\ 1211-Geneva,\\Switzerland\\
E-mail: Daniel.Ferenc@CERN.ch} 


\maketitle\abstracts{ 
New developments in
HPD design are presented, triggered by applications
in high energy physics and astrophysics.
The presented HPD designs are based on three innovations.
(i)
In order to achieve the highest possible surface coverage in a
RICH detector,
we introduced a photoelectron focussing
method which is efficient to
the periphery of the photocathode.
(ii)
To prevent positive ion feedback in HPDs,
we introduced a 
permanent potential barrier in front of the anode.
(iii)
To replace a transmittive by a reflective photocathode,
we arrived at a conceptually new
HPD design with surprisingly good imaging characteristics,
high quantum efficiency and low cost.
}

\section{Introduction}
With the onset of new technologies, 
Hybrid Photon Detectors (HPDs) became the
most favourable option for detection of Cherenkov photons
in large area Ring Imaging Cherenkov (RICH)
detectors.
Modern HPD detectors comprise high quantum efficiency,
high photoelectron collection efficiency
and sharp
image reproduction.
We present some new developments in
HPDs, of particular importance for 
applications in RICH detectors. The goals achieved in the
presented HPD designs are:
\begin{enumerate}
\item
minimized dead area of individual HPDs,
and consequently maximized active area of
a RICH detector, 81\% in a hexagonal HPD packing,
\item
protection against the positive ion feedback, particularly
important in gamma ray astronomy, and
\item
application of a low cost and high quantum efficiency 
reflective photocathode in an HPD.
\end{enumerate}
For all the electron optics simulations presented,
the SIMION 3D software~\cite{SIMION} has been used.
In the figures presented, only functional elements (conductors)
are shown. Electrons are simulated with the initial energy of 0.25 eV
and emission angles
+45$^{\circ}$, -45$^{\circ}$ and 0$^{\circ}$ relative to the
normal.

\section{``Killing the dead area"}
The most important problem in the integration of HPDs into
a matrix of a large-surface RICH detector is the low overall
photon-sensitive surface coverage, caused by a typically high 
HPD dead area. HPDs have usually been designed as stand-alone
devices, and very little, if any care has been taken of the relationship
between the physical and the sensitive surface areas.
In those applications when the Cherenkov photon detection pixel size
of 1-2 cm is sufficiently small, one can use single-pixel HPDs 
(without internal imaging) and take care of the large dead area by
focussing the light to the sentive area by the means of
lenses or
Winstone cones. In the applications which require a smaller pixel
size (e.g. 1 mm), one has to use large
diameter HPDs with internal imaging.

\begin{figure}[hbp]
\epsfig{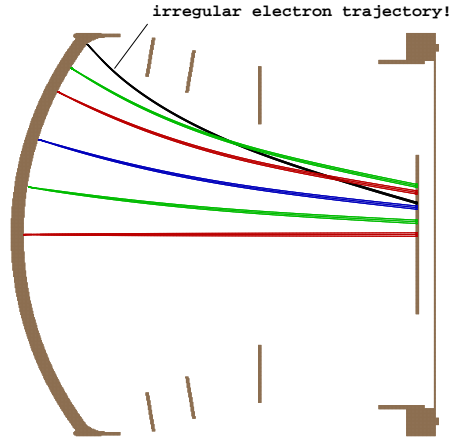}
\caption{
\noindent
Proximity-focussing 5 inch diameter pad-HPD.
Photons enter HPD from the left side,
photoelectrons are
focussed onto
the silicon-pad
detector on the right side.
Photoelectrons emerging from the periphery 
of the photocathode are incorrectly focussed.
Electrodes are kept at the following potentials, from left to
right, respectively: -20 kV, -15 kV, -11 kV, -4.7 kV and 0 V.
Electrons are simulated with an initial energy of 0.25 eV
and an emission angle of
+45$^{\circ}$, -45$^{\circ}$ and 0$^{\circ}$ relative to the
normal.
}
\label{FPP01}
\end{figure}

\begin{figure}[htp]
\epsfig{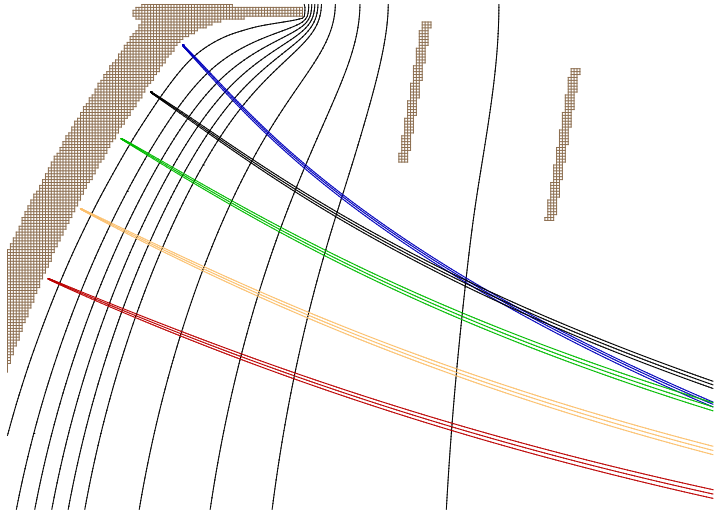}
\caption{
\noindent
Incorrect electron focussing close to the 
periphery of the photocathode is due to the
small radius of curvature of equipotential lines. 
}
\label{FPP02}
\end{figure}

\begin{figure}[hbp]
\epsfig{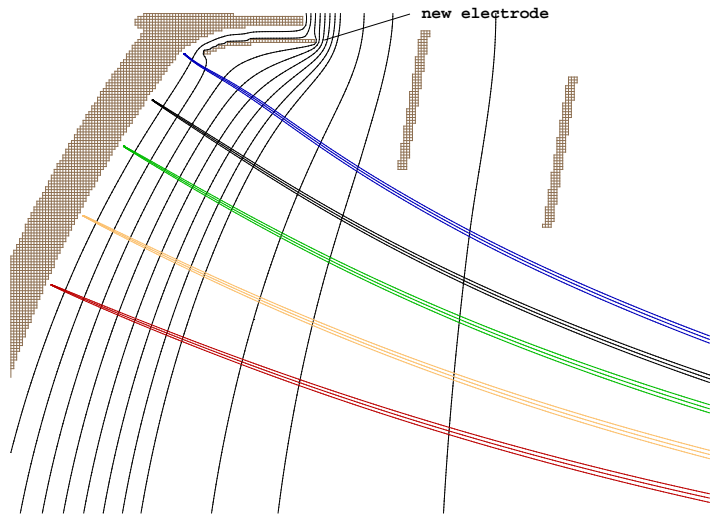}
\caption{
\noindent
The new electrode
allows the
potential lines to be conducted out from the HPD through
the created slot, curing 
the unwanted
strong curvature of the equipotential lines seen in Fig.~\ref{FPP03}.
}
\label{FPP03}
\end{figure}

Proximity-focussing HPD designed for
the LHCb experiment~\cite{LHC-B-TP}, is an example of a
large imaging HPD 
(Fig.~\ref{FPP01}). It has been conceptually designed 
to have a small dead area, for close hexagonal packing. 

Photons enter the HPD detector
from the left side, and
photoelectrons (emerging from the photocathode on the 
internal surface of the entrance window) undergo
acceleration and (electrostatic) focalization
onto the silicon-pad
detector on the right side. There is a good mapping
between the image on the photocathode and the 
projected image on the pad detector, except 
for the region close to the periphery of the photocathode.
Although the physical shape of this HPD is 
optimized for close packing (the photocathode has
been extended towards the periphery as much as possible),
there is still a rather large {\it functionally} dead area.
\begin{figure}[htbp]
\epsfig{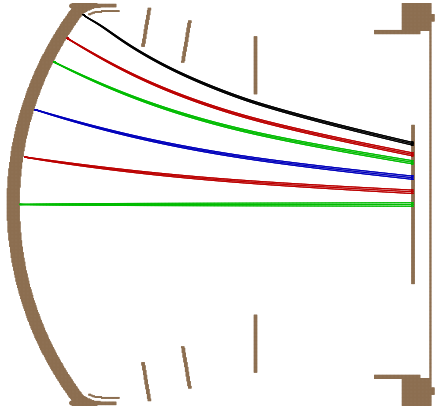}
\caption{
\noindent
An appropriate focussing of all electrons, including
those from the edge of the photocathode, results from 
the application of a new electrode. Potential lines are ``conducted"
out from the tube through the slot between this electrode and the 
window support.
Electrodes are kept at the following potentials, from left to
right, respectively: -20 kV, -19.45 kV, -15 kV, -11 kV, -4.7 kV and 0 V.
}
\label{FPP04}
\end{figure}

The reason for the failure 
becomes evident from Fig.~\ref{FPP02}.
Equipotential lines 
have a rather small radius of curvature 
close to the edge of the photocathode.
Photoelectrons emitted
from that region are therefore too strongly accelerated 
along the potential gradient, i.e.
towards
the center of the HPD.

To fix this problem, one should reduce the curvature
of the potential distribution.
One is tempted to redesign the window supporting
structure and let potential lines leave the tube,
but for constructional reasons (related in fact to the
maximal exclusion of the mechanical dead area),
this was not possible.
Therefore we searched for another solution
with the basic idea to reduce the field curvature
by ``conducting" some of the potential lines 
out from the 
tube, around the metallic window support.
The solution
was found in the creation of a slot which 
acts as a ``potential-conductor", see Fig.~\ref{FPP03}.
The slot was created by the insertion of a specially shaped
new electrode.
The unwanted potential lines are indeed conducted away
through the slot between the new electrode 
and the body of the tube, and
the resulting field
in the problematic peripheral region has evidently lost its
strong curvature, 
see Fig.~\ref{FPP03}, Fig.~\ref{FPP04}.
\begin{figure}[htbp]
\epsfig{file=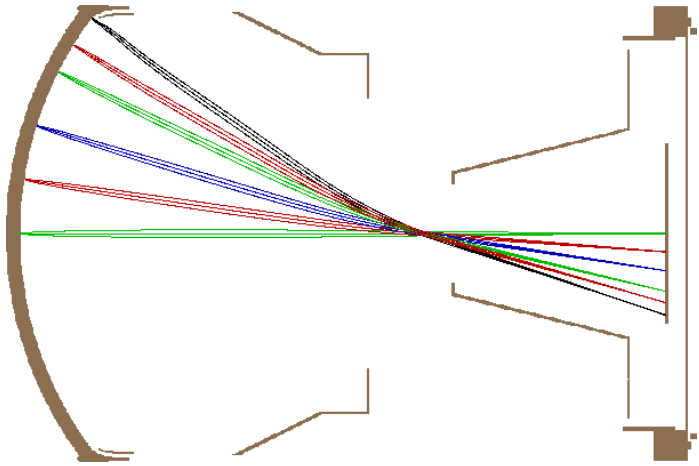,bbllx=0,bblly=150,bburx=792,bbury=620,width=12cm}
\caption{
\noindent
A cross-focussing 5-inch diameter HPD, with superior imaging characteristics.
Electrodes are kept at the following potentials, from left to
right, respectively: -20 kV, -19.97 kV, -19.4 kV, +100 V, and 0 V.
}
\label{FPP1}
\end{figure}

The same method has been successfully applied in a different design, the
so called
cross-focussing HPD design, shown
in Fig.~\ref{FPP1}. This HPD can also be hexagonally packed with
the same surface coverage of
81\% .

Apart from providing a much narrower spread
of photoelectrons on the silicon pad detector, and thus a superior 
imaging performance, this design also provides a simple way to apply
the protection against the positive ion feedback~\cite{NIM},
which is the subject of the following section.

\section{Potential barrier - protection against the ion feedback}

\begin{figure}[htbp]
\epsfig{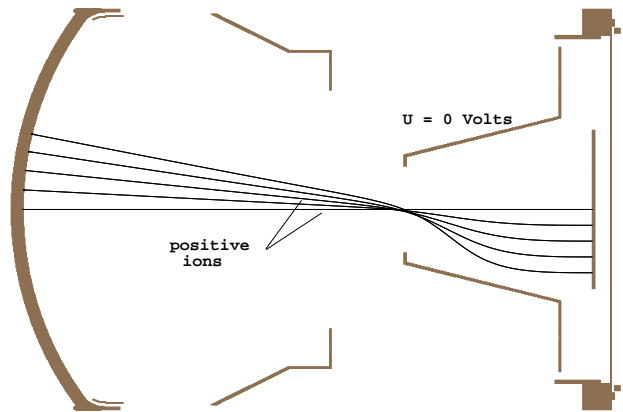}
\caption{
\noindent
Positive ion trajectories, when the conical electrode is set
to the anode potential (0 V). Ions
emerge from the surface of the anode (right) and become accelerated
towards the photocathode (left), eventually producing damage
and operational noise.
}
\label{FPP2}
\end{figure}
\begin{figure}[htbp]
\epsfig{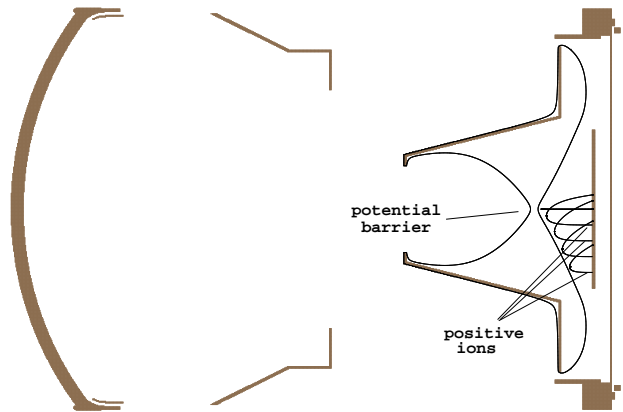}
\caption{
\noindent
HPD design with a conical barrier--electrode at
+100 V. Between the barrier--electrode
and the anode a potential barrier is established (see Fig.~\ref{FPP4}) to
repel back the positive ions emerging from the
anode surface.
}
\label{FPP3}
\end{figure}

Air Cherenkov Telescopes (ACT) have been considered
the ultimate instruments
for the ground based detection of
high energy cosmic gamma rays~\cite{ACT-general,Norbert}.
In order to lower the
energy threshold for the detection of cosmic gamma rays
down to 20 GeV -- to explore the only unexplored
window in cosmic electromagnetic spectrum (20 GeV to 300 GeV) --
one should both
increase the detector area, and achieve an unprecedented
photon detection with single photon
sensitivity and very high efficiency. Considering photon sensors,
HPDs currently present
the most promising solution.
However,
commercial devices have still some
serious drawbacks and need further improvement.
In particular, it is very important to
reduce the internal instrumental noise below the present limits,
 because other
sources of noise in imaging
air Cherenkov detectors (like the night sky background)
are irreducible.

The presence of positive ions in a vacuum
tube is particularly devastating because
the acceleration and subsequent
dumping of positive ions into a
photocathode leads both
to creation of noise
through electrons released,
and
to a damage of the
photocathode
~\cite{Razmik,Eckart-NIM}.
In the high-vacuum tubes the vast majority of
positive ions do not originate from residual gas, but
rather from the impact of
accelerated photoelectrons
on the surface of the anode.
Cesium ions are particularly
abundant because they usually
spread inside tubes during
and after the
manufacturing of photocathodes.

Trajectories of singly charged positive
ions are shown in
Fig.~\ref{FPP2}, 
emerging from the anode at normal incidence
with energy 44 eV.
Note that the angular and
energetic distributions of positive ions
are at this point unknown. We have worked out a
measurement scheme, but
at the time being 
we are using only a very rough
estimate that the ions could reach an energy of about 30 eV.

It has been previously demonstrated
~\cite{NIM} 
that the
insertion of an electrostatic potential barrier
close to the anode
solves the ion feedback problem.
Apart from being complete, this
solution is easy to
implement and it preserves cylindrical symmetry of the device.

\begin{figure}[htbp]
\epsfig{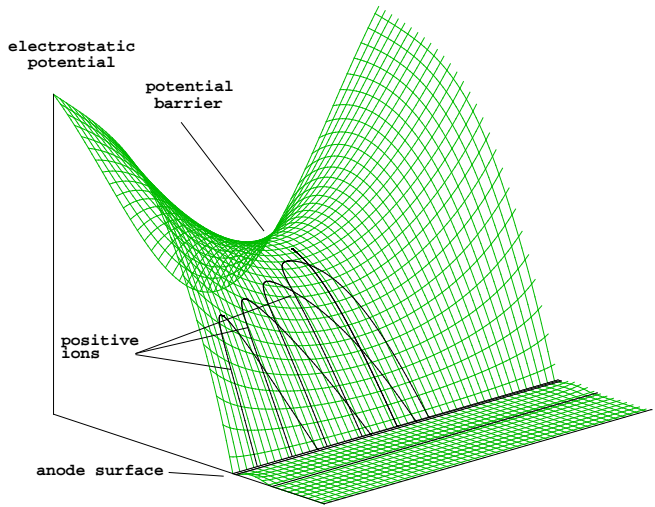}
\caption{
\noindent
Potential distribution in front of the
anode plane of HPD from
Fig.~\ref{FPP3}.
Positive ions of energy E$_{ion}$=44 eV
and emission angle normal to the anode surface
start ``climbing" the potential barrier (E$_{b}$=45 eV)
but eventually become repelled.
}
\label{FPP4}
\end{figure}

As demonstrated in 
Fig.~\ref{FPP3},
the functionality of the conical barrier--electrode
is simple:
being kept at a potential somewhat higher
than the anode potential, it breaks down the monotonous 
decrease of the potential for positive particles towards the 
photocathode, and
creates a potential barrier
in front of the anode. The barrier
prevents positive 
ions from penetrating further towards the photocathode.
The potential distribution in front of the anode plane 
is shown in a magnified view in Fig.~\ref{FPP4}.
Trajectories of singly charged positive
ions are simulated with identical initial conditions
like before.

The precision
of the potential on the barrier--electrode,
required
for stable electron focussing, is
not a critical issue
-- variations of even 10\% on the potential will leave
the electron focussing essentially
unchanged~\cite{NIM,Diplomski}.
The most common
voltage supply may be therefore used
to bias the barrier--electrode, while a separate,
unipolar and
very stable
voltage supply could be
used to bias the focussing
electrodes.

\section{HPD with a reflective photocathode}

Semi-transparent photocathodes, commonly
used in photon detectors, present a problem {\it per se}:
they need to be opaque for photons, but at the same time
transmittive for
photoelectrons. 

\begin{figure}[htp]
\epsfig{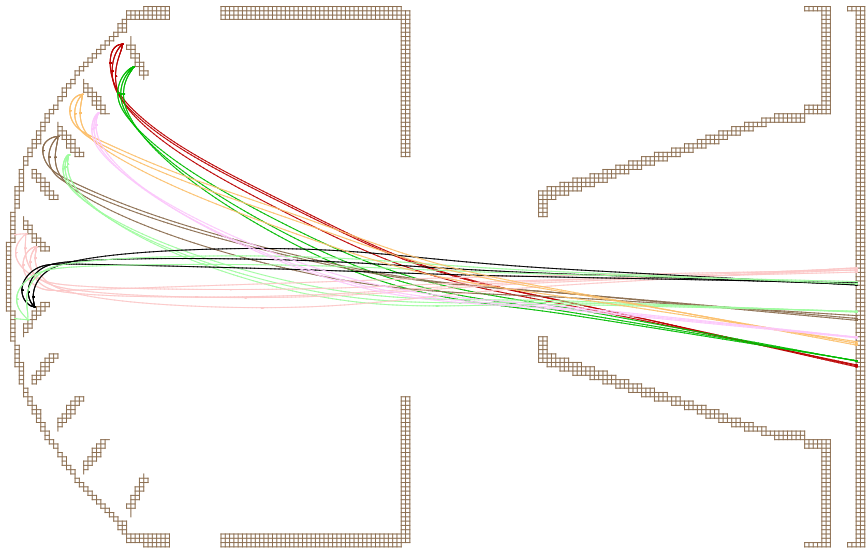}
\caption{
\noindent
Imaging
HPD with reflective photocathode. Photoelectrons emerge from the 
photocathode attached to the surface of conical ``blinds".
Note the surprisingly good imaging performance!
Electrodes are kept at the following potentials, from left to
right, respectively: -20 kV, -19.4 kV, +100 V, and 0 V.
}
\label{FPP6}
\end{figure}

An alternative solution is to use a photocathode in a reflective 
mode, i.e. in a configuration when photoelectrons
emerge from the same surface through which 
photons enter.
A considerably higher quantum
efficiency is granted, but perhaps equally important,
the photocathode manufacturing process is not any more
strictly constrained to extremely high tolerances. 
In particular,
there is no need to perform some of the most complicated
stages in the processing of the III-V photocathodes
(like e.g. GaAsP), namely the attachment of 
the epitaxially grown surface
onto the entrance window of the phototube,
and then the removal (usually by etching) of the substrate
from the opposite side. 

Motivated by these considerations, we have developed 
a conceptually new HPD device - the imaging
HPD with reflective photocathode, see 
Fig.~\ref{FPP6}. This cylindrically symmetric device
converts photons into photoelectrons in the photocathodes
mounted on the 
surface of the conical ``blinds", 
attached mechanically
and electrically to the 
entrance window.
After a detailed electron optics optimization,
this device provided a surprisingly
good imaging quality, see Fig.~\ref{FPP6}.
Note that two imaging operational modes are
possible: (i) a mapping of each individual
blind electrode into a single ``point",
in which case the silicon detector surface should
be placed slightly closer towards the photocathode
than in Fig.~\ref{FPP6}, and (ii) a point-to-point
mapping, as shown in Fig.~\ref{FPP6}.

Among the drawbacks of this particular
design, one should note a compromised photon
angular acceptance and a relatively large
difference in the time of flight
of photoelectrons emitted from different
points on the same blind.

\section*{Acknowledgments}
I would like to thank Eckart Lorenz
for introducing me into the world of HPDs.
The LHCb-related applications presented in this paper
were done in a collaboration with Jacques Seguinot and
Tom Ypsilantis, and the concept of reflective photocathode
HPDs has been innitiated in a collaboration with Guy Pai\' c.
Many thanks to Dario Hrupec who participated
in part of the presented activities.

\end{document}